\documentclass[prl,twocolumn,superscriptaddress,amsmath,amssymb]{revtex4}
\usepackage{graphicx}% Include figure files
\usepackage{dcolumn}% Align table columns on decimal point
\usepackage{bm}% bold math
\usepackage{color}% perhaps for xfig exports needed
\usepackage{epsfig}%perhaps necessary for xfig exports
\usepackage{amssymb}

\begin{document}

\newcommand{\commute}[2]{\left[#1,#2\right]}
\newcommand{\bra}[1]{\left\langle #1\right|}
\newcommand{\ket}[1]{\left|#1\right\rangle }
\newcommand{\anticommute}[2]{\left\{  #1,#2\right\}  }
\renewcommand{\arraystretch}{2}

\title{Universal phase shift and non-exponential decay of driven single-spin oscillations}

\author{F.H.L. Koppens}
\affiliation{Kavli Institute of NanoScience Delft, P.O. Box 5046, 2600 GA Delft, The Netherlands}
\author{D. Klauser}
\affiliation{Department of Physics and Astronomy, University of Basel, Klingelbergstrasse 82, CH-4056 Basel, Switzerland}
\author{W. A. Coish}
\affiliation{Department of Physics and Astronomy, University of Basel, Klingelbergstrasse 82, CH-4056 Basel, Switzerland}
\author{K. C. Nowack}
\affiliation{Kavli Institute of NanoScience Delft, P.O. Box 5046, 2600 GA Delft, The Netherlands}
\author{L.P. Kouwenhoven}
\affiliation{Kavli Institute of NanoScience Delft, P.O. Box 5046, 2600 GA Delft, The Netherlands}
\author{D. Loss}
\affiliation{Department of Physics and Astronomy, University of Basel, Klingelbergstrasse 82, CH-4056 Basel, Switzerland}
\author{L.M.K. Vandersypen}
\affiliation{Kavli Institute of NanoScience Delft, P.O. Box 5046, 2600 GA Delft, The Netherlands}

\begin{abstract}
We study, both theoretically and experimentally, driven Rabi oscillations of a single electron spin coupled to a nuclear spin
bath. Due to the long correlation time of the bath, two unusual features are observed in the oscillations. The decay follows a
power law, and the oscillations are shifted in phase by a universal value of $\sim\pi/4$. These properties are well
understood from a theoretical expression that we derive here in the static limit for the nuclear bath. This improved
understanding of the coupled electron-nuclear system is important for future experiments using the electron spin as a qubit.
\end{abstract}

\maketitle 
A quantum bit is engineered such that its coupling to the disturbing environment is minimized. Understanding and
controlling this coupling is therefore a major subject in the field of quantum information processing. It is not solely the
coupling strength but also the dynamics of the environment that governs the quantum coherence. In particular, the limit where
these dynamics are slow compared to the evolution of the quantum system is interesting. The well-known Markovian Bloch equations
that describe the dynamics of a driven system, including the exponential decay of the longitudinal and transverse magnetization
\cite{bloch1946}, then lose their validity. Such deviations from the exponential behavior have been studied
theoretically \cite{lossprb05,taylor06} and experimentally, for instance in superconducting qubit systems \cite{ithier05}.

An electron spin confined in the solid state is affected predominantly by phonons via the spin-orbit interaction
\cite{khaetskiinazarov,golovach03,elzerman,kroutvar,amasha-2006}, and by nuclear spins in the host material via the hyperfine interaction. At low
temperature, coupling to the nuclear spins is the dominant decoherence source \cite{khaetskii02,merkulov,coish04,coish05,johnsonnature,koppensscience,petta,laird06}. Although this strong coupling leads to an apparent
decoherence time $T_2^*$ of the order of 20 ns when time-averaged over experimental runs, the decoherence time $T_2$ strongly depends on the dynamics in the nuclear spin bath. This
typical nuclear spin dynamics is very slow, because the nuclear spins are only weakly coupled with each other and the bath itself
is coupled very weakly to its dissipative environment (like phonons). This implies that here, the Markovian Bloch equations
are not valid.

Here we study the dynamics and decoherence of an electron spin in a quantum dot that is coherently driven via pulsed magnetic
resonance, and is coupled to a nuclear spin bath with a long correlation time. We find experimentally that, remarkably, the
electron spin oscillates coherently, even when the Rabi period is much longer than $T_2^*=10-20$ ns. In addition, the
characteristics of the driven electron spin dynamics are unusual. The decay of the Rabi oscillations is not exponential but
follows a power law and a universal (parameter independent) phase shift emerges.
%A power-law decay is generally difficult to observe because it usually appears at long timescales. Here, the power-law decay is
%already valid after a short time (see below), allowing it to be observed experimentally. In this work, we present the observation
%of a power-law decay as well as a universal phase shift.
We compare these experimental results with a theoretical expression, derived in the limit of a static nuclear spin bath.

We consider a double quantum dot with one electron in each dot and a static external magnetic field in the z-direction, resulting
in a Zeeman splitting $\epsilon_z=g \mu_B B_z$. The spin transitions are driven by a burst of a transverse oscillating field along the $x$-direction with amplitude $B_{\mathrm{ac}}$ and frequency $\omega$, which is generated by a current $I_\mathrm{s}$ through a
microfabricated wire close to the double dot \cite{koppensnature}. The interaction between the electron spin and the nuclear bath
is described by the Fermi contact hyperfine interaction $\vec{S}\cdot\vec{h}$, where $\vec{h}$ is the field generated by the
nuclear spins at the position of the electron. For a large but finite number of nuclear spins ($N \sim 10^6$ for lateral GaAs
dots) $h_{z}$ is Gaussian distributed (due to the central-limit theorem) with mean $h_0=\overline{h_{z}}$ and variance
$\sigma^2=\overline{(h_{z}-h_{0})^2}$ \cite{khaetskii02,merkulov,coish04}. For a sufficiently large external magnetic field
($\epsilon_z\gg\sigma$), we may neglect the transverse terms $S_{\bot}\cdot h_{\bot}$ of the hyperfine interaction that give rise
to electron-nuclear-spin flip-flops (see below). Furthermore, if the singlet-triplet energy splitting $J$ is much smaller than
both $\epsilon_z$ and $g \mu_B B_{\mathrm{ac}}$, we may treat the spin dynamics of the electrons in each dot independently (valid
for times less than $1/J$). 

For each dot we thus have the following spin Hamiltonian ($\hbar=1$):
\begin{equation}\label{eq:hamiltonian}
H(t)=\frac{1}{2}(\epsilon_z+h_z)\sigma_z+\frac{b}{2}\cos(\omega t)\sigma_x,
\end{equation}
where $\sigma_i$ (with $i=x, z$) are the Pauli matrices and $b=g \mu_B B_{\mathrm{ac}}$ (taken to be equal in both dots).
%We note two important points about this Hamiltonian. First, it contains only one decoherence source: the uncertainty of the
longitudinal nuclear field $h_z$. Here, $h_z$ is considered as completely static during the electron spin time evolution. This
is justified because the correlation time of the fluctuations in the nuclear-spin system due to dipole-dipole and
hyperfine-mediated interaction between the nuclear spins, which is predicted to be $\gtrsim 10-100\,\mu$s
\cite{coish04,khaetskii02,merkulov,yao,sousa03,klauser06,deng,dengerratum}, is much larger than the timescale for electron
spin dynamics considered here (up to $1\mu$s).
%This is because the correlation time of the fluctuations in $h_z$ is long and thus the spectral density $S_h(\omega)$ of the nuclear field has a small cutoff frequency $\omega_c$ (smaller than 1 MHz. according to several theories \cite{sousa,dutt,coish}). For a driven evolution (with rate $b/2$) only two contributions from the spectral density can contribute to dephasing \cite{slichter,geva}: $S(\omega=0)$ and $S(\omega=b/2)$. This implies that for $\omega_c<b/2$ we have to consider only the zero-frequency components $S(\omega=0)$ ($h_z$ static). We will see below, where theoretical results are compared with the experimental data, that this approximation is valid for strong enough driving fields.

In the experiment, the electron spin state is detected in a regime where electron transport through the double quantum dot occurs
via transitions from spin states with one electron in each dot (denoted as (1,1)) to the singlet state $\ket{S(0,2)}$ with two
electrons in the right dot. These transitions, governed via the tunnel coupling $t_c$ by the tunneling Hamiltonian $H_{t_c}={t_c}
\ket{S(1,1)} \bra{S(0,2)}+H.c.$, are only possible for anti-parallel spins, because
$\bra{\uparrow\uparrow}H_{t_c}\ket{\mathrm{S(0,2)}}=
  \bra{\downarrow\downarrow}H_{t_c}\ket{\mathrm{S(0,2)}}=0$, while
$\bra{\downarrow\uparrow}H_{t_c}\ket{\mathrm{S(0,2)}},
  \bra{\uparrow\downarrow}H_{t_c}\ket{\mathrm{S(0,2)}}\neq0$. Therefore, the states with even spin parity (parallel spins) block transport, while the states with odd spin parity (antiparallel spins) allow for transport.
If the system is initialized to an even spin-parity state, the oscillating transverse magnetic field (if on resonance) rotates
one (or both) of the two spins and thus lifts the blockade \cite{koppensnature}. Initializing to $\ket{\uparrow}$ in both dots
(the case with $\ket{\downarrow}$ gives the same result), we calculate the probability for an odd spin parity $P_{odd}$ under
time evolution for each of the two spins governed by the Hamiltonian in Eq.(\ref{eq:hamiltonian}).
%In this calculation we take into account that the nuclear field $h_z$ is static during the spin evolution but changes on the timescale of a measurement, %which is an average over 15 s for each value of the burst time $t$ (spin evolution time).

Introducing the detuning from resonance $\delta_{\omega}=\epsilon_z
+h_z-\omega$,
the probability to find spin up for a single value of $h_z$ in the rotating
wave approximation (which is valid for $(b/\epsilon_z)^2\ll1)$ is given by
\begin{equation} P_{\uparrow,\delta_{\omega}}(t)=\frac{1}{2}\left[1+\frac{4\delta_{\omega}^2}{b^2+4\delta_{\omega}^2}
+\frac{b^2}{b^2+4\delta_{\omega}^2}\cos\left(\frac{t}{2}\sqrt{b^2+4\delta_{\omega}^2}\right)\right].
\end{equation}
Assuming that $\omega=h_0+\epsilon_z$,  i.e., $\delta_{\omega}=h_z-h_0$, we find
when averaging over the Gaussian distribution of $h_z$ values (see \cite{EPAPS})
\begin{eqnarray}\label{eq:integral}
 P_{\uparrow}(t)\sim
  \frac{1}{2}+C+\sqrt{\frac{b}{8\sigma^2 t}}\cos\left(\frac{b}{2}t+\frac{\pi}{4}\right)+\mathcal{O}\left(\frac{1}{t^{3/2}}\right),
\end{eqnarray}
for $t\gg \max({\frac{1}{\sigma},1/b, b/2\sigma^2})$, with $C=\frac{1}{2}-\frac{\sqrt{2 \pi} b}{8
\sigma}\exp\left(\frac{b^2}{8\sigma^2}\right)\mathrm{erfc}\left(\frac{b}{2\sqrt{2}\sigma}\right).$
We can now calculate the probability of finding an odd spin-parity state
taking $\omega=h_0+\epsilon_z$ for both dots and drawing the value of $h_z$
independently from a distribution with width $\sigma$ in each dot:
%  if we assume that in both dots the mean of the Gaussian distribution is zero (no nuclear polarization) and their widths are equal:
\begin{eqnarray}\label{eq:poddtwodot}
P_{\mathrm{odd}}(t) &=& P_{\uparrow,L}(t) (1-P_{\uparrow,R}(t)) +(1-P_{\uparrow,L}(t))P_{\uparrow,R}(t) \notag
\\
&=& \frac{1}{2}-2C^2-C \frac{f(t)}{\sqrt{t}}-\frac{g(t)}{t}+\mathcal{O}\left(\frac{1}{t^{3/2}}\right);\\
f(t)&=&\sqrt{\frac{2 b}{\sigma^2}}\cos\left(\frac{b
    t}{2}+\frac{\pi}{4}\right), \label{ft}\\
g(t)&=&\frac{b}{8 \sigma^2}\left[1+\cos\left(b
    t+\frac{\pi}{2}\right)\right].
\end{eqnarray}
This result is valid for times $t \gtrsim \max(1/\sigma,1/b,b/2\sigma^2)\sim$ 20ns for a 1.4 mT nuclear field (see below) and $b\leq2\sigma$ (accessible experimental regime). The $1/t$-term oscillates with the double Rabi frequency which is the result of both spins being rotated simultaneously (see also \cite{koppensnature}).
This term only becomes important for $b
>\sigma$, because in that case for both spins most of the nuclear-spin distribution is within the Lorentzian lineshape of the Rabi resonance. The
$1/\sqrt{t}$-term oscillates with the Rabi frequency and originates from only one of the two spins being rotated
\cite{koppensnature}. This term is important when $b<\sigma$, i.e., when only a small fraction of the nuclear-spin distribution
is within the lineshape of the Rabi resonance.

\begin{figure}[b]
          \includegraphics[scale=0.55]{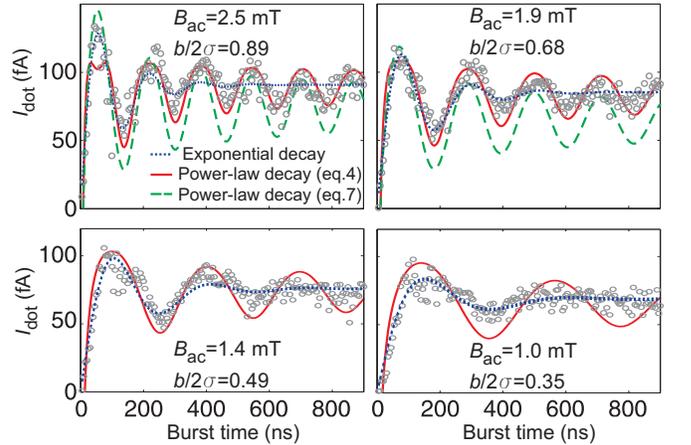}
     \caption{(Color online) Rabi oscillations for four different driving fields $B_{\mathrm{ac}}$ ($B_{z}=$55 mT, $g$=0.355 and $\sigma=g\mu_B$(1.4 mT)).
     %The data were obtained by means of the pulse technique as described in Ref. \cite{koppensnature}.
The gray circles represent the experimentally measured dot current (averaged over 15 s for each value of $t$), which reflects the
probability to find an odd spin-parity state after the RF burst that generates $B_{\mathrm{ac}}$. The dotted, solid and dashed lines
represent the best fit to the data of an exponentially decaying cosine function and the derived analytical expressions for
$P_{odd}(t)$ and $P_{odd}^{(1)}(t)$ (Eqs. (\ref{eq:poddtwodot}) and (\ref{eq:poddonedot})) respectively. For clarity, the dashed
line is shown only for the top two panels. The fit was carried out for the range 60 to 900 ns and the displayed values for
$B_{\mathrm{ac}}$ were obtained from the fit with $P_{odd}(t)$ (Eq. (\ref{eq:poddtwodot})). We fit the data with an exponentially
decaying cosine with a tunable phase shift that is zero at $t=0$: $a_1 e^{-t/a_2} [\cos(\phi)-\cos(2\pi t/a_3+\phi)]
+a_4(1-e^{-t/a_2})$. The last term was added such that the saturation value is a fit parameter
as well. We note that the fit is best for $\phi=\pi/4$, as discussed in the text.}
     \label{oscillations}
\end{figure}

We also give the expression for $P_{odd}(t)$ for the case where only one of the two spins is on resonance
($\epsilon_z+h_{0}-\omega=0$), while the other is far off-resonance ($|\epsilon_z+h_{0}-\omega|\gg \sigma$). In this case the
spin in one dot always remains up while the spin in the other dot rotates. This leads to
\begin{eqnarray}\label{eq:poddonedot}
P^{(1)}_{\mathrm{odd}}(t)= 1-P_{\uparrow}(t) = \frac{1}{2}-C-\frac{f(t)}{4 \sqrt{t}}+\mathcal{O}\left(\frac{1}{t^{3/2}}\right),
\end{eqnarray}
with the same range of validity as in Eq.(\ref{eq:poddtwodot}). We see that the $1/t$-term, which oscillates with frequency $b$,
is not present in this case.

The expressions for $P_{odd}(t)$ (Eqs. (\ref{eq:poddtwodot}) and (\ref{eq:poddonedot})) reveal two interesting features: the
\emph{power-law decay} and a universal \emph{phase shift} of $\pi/4$ (see Eq. (\ref{ft})) in the oscillations which is independent
of all parameters. These features can both only appear if the nuclear field $h_z$ is static during a time much longer than the
Rabi period. This is crucial because only then the driven spin coherence for one fixed value of $h_z$ is fully preserved. Because
different values of $h_z$ give different oscillation frequencies, the decay is due to averaging over the distribution in $h_z$.

The phase shift is closely related to the power-law decay because it also finds its origin in the off-resonant contributions.
These contributions have a higher Rabi frequency and shift the average oscillation in phase. This universal phase shift therefore
also characterizes the spin decay, together with the power law. Interestingly, the specific shape of the distribution in $h_z$
(as long as it is peaked around the resonance) is not crucial for the appearance of both the power-law decay and the phase shift \cite{EPAPS}.
The values of the decay power and the phase shift are determined by the dependence of the oscillation frequency on $h_z$ (in this
case $\sqrt{b^2+4\delta_\omega^2}$).

A power-law decay has previously been found theoretically in \cite{khaetskii02,coish04,harmon2003,huang:2005a} and both a
power-law decay ($1/t^{3/2}$) and a universal phase shift also appear in double dot correlation functions \cite{coish05,klauser06}. In
\cite{laird06} a singlet-triplet correlation function was measured, but the amplitude of the oscillations was too small for the phase shift
and the power-law decay to be determined. Here, we consider driven Rabi oscillations of a single electron spin with a power-law
decay of $1/\sqrt{t}$ that is already valid after a short time $1/\sigma\sim20$ ns. Therefore, the amplitude of the driven spin
oscillations is still high when the power-law behavior sets in, even for small driving fields ($b<2\sigma$) which are experimentally easier to achieve. The power-law decay and the phase shift thus should be observable in the experiment.

%The fact that the power-law decay and phase shift are also present for these driven Rabi oscillations in a measurable quantity $P_{\mathrm{odd}}(t)$ makes %them also directly accessible to the experiment.

\begin{figure}[t]
          \includegraphics[scale=0.55]{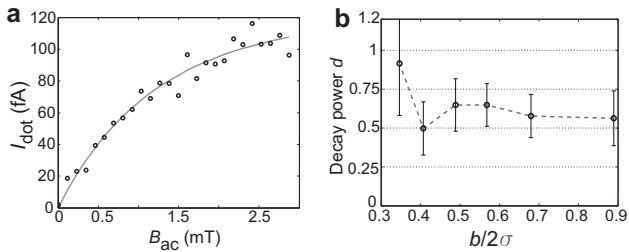}
\caption{ a) Dot current after an RF burst of 950 ns as a function of $B_{\mathrm{ac}}$, approximately representing the steady-state value.
The solid curve is the best fit with $a_1(\frac{1}{2}-2C^2)$: the steady state expression of Eq. (\ref{eq:poddtwodot}) with $a_1$
and $\sigma$ as fit parameters. We find, for the 95\%-confidence interval, $\sigma=g\mu_B(1.0-1.7$ mT). b) Decay power obtained
from the best fit of the data (partially shown in Fig. \ref{oscillations}) with the expression $a_1+a_2\cos(2\pi t
/a_3+\pi/4)/t^{d}$, where $a_{1,2,3}$ and $d$ are fit parameters. }
     \label{steadystate}
\end{figure}

We now discuss the observation of the power-law decay in the experimental data of which a selection is shown in Fig.
\ref{oscillations}. The data are obtained with the same device and under the same experimental conditions as in
\cite{koppensnature}. A fit is carried out to the observed oscillations for four different driving fields $B_{\mathrm{ac}}$ (Fig. \ref{oscillations}), with three different fit functions: the theoretical expressions (Eqs. (\ref{eq:poddtwodot}) and
(\ref{eq:poddonedot}) with $b$ and a constant scaling factor as fit parameters) and an exponentially decaying cosine. The width
of the nuclear distribution $\sigma=g\mu_B(1.4$ mT) is obtained from a fit of the steady state value $\frac{1}{2}-2C^2$ of $P_{odd}(t)$
to a dataset obtained at $t=$ 950 ns (Fig. \ref{steadystate}a).

For the range $B_{\mathrm{ac}}\geq$ 1.9 mT, we find good agreement with the model that predicts a power-lay decay of $1/\sqrt{t}$ (Eq. (\ref{eq:poddtwodot}); $h_0$ equal for both dots), while the fit with an exponentially decaying cosine is poor (blue lines in Fig.
\ref{oscillations}). The power of the decay is independently verified by means of a fit to the data with $a_1+a_2\cos(2\pi t
/a_3+\pi/4)/t^{d}$ where, besides $a_{1,2,3}$, the power $d$ of the time $t$ is a fit parameter as well. We find values of
$d\sim0.6$ (Fig. \ref{steadystate}b), close to the predicted $1/\sqrt{t}$-dependence.

We see much better correspondence of the data with Eq. (\ref{eq:poddtwodot}) than with Eq. (\ref{eq:poddonedot}), from which we can conclude that the mean of the Gaussian distribution $h_0$ is comparable for both dots (in equilibrium, we expect $h_0 \sim0 $ in both dots). There might however still be a small difference in $h_0$ between the two dots, which we cannot determine quantitatively because the two models describe only two limiting cases. If present, such a difference in $h_0$ could help explain the small deviation between data and model at the first oscillation for $B_{\mathrm{ac}}=2.5$ mT. It could originate from asymmetric feedback of the electron spins on the respective nuclear spin baths, e.g. due to unequal dot sizes, leading to different hyperfine coupling constants. 

Another observation is that for small driving fields, $B_{\mathrm{ac}}<1.9$ mT, we see that the damping is faster than predicted. Possible explanations for this effect are corrections due to electron-nuclear flip-flops (transverse terms in  the hyperfine Hamiltonian) or electric field fluctuations. Electron-nuclear flip-flops may become relevant on a timescale $\sim{\epsilon_z}/\sigma^2\sim1\,\mu s$ in this experiment.  Electric field fluctuations can couple to spin states via the spin-orbit interaction \cite{borhani06} or a finite electric-field dependent exchange coupling.

%are the corrections due to electron-nuclear flip-flops (transverse terms of the hyperfine Hamiltonian)
%which have a timescale of ${\epsilon_z}/\sigma^2\sim750$ ns \cite{coish05} (for $B_z$=55 mT and $\sigma=1.5$ mT). These
%corrections have more effect if the Rabi period is comparable to or smaller than this timescale, in other words if
%$1/b\leq{\epsilon_z}/\sigma^2$. A second possible explanation is that, at a time $\sim1/J$, the exchange interaction becomes
%important and fluctuations in $J$ due to electric field fluctuations cause additional decoherence \cite{coish05}.
%the contribution from the so far neglected Rabi frequency component of the nuclear bath noise power spectrum $S(\omega_{rabi})$. This contribution - which %can give rise to a faster and different type of decay - is important if the Rabi time exceeds the correlation time of the nuclear bath. The observed %transition to a faster decay would imply a correlation time of $\sim 0.5\mu s$ which is a realistic value regarding the existing theories %\cite{dutt,coish,sousa,taylor2005}.

\begin{figure}[t]
          \includegraphics[scale=0.78]{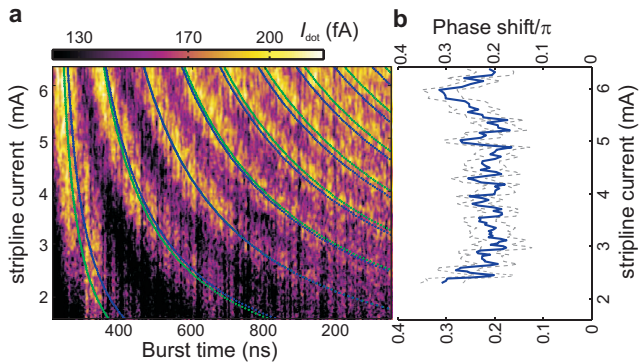}
   %  \centering
     \caption{ a) The dot current (represented in colorscale) is displayed over a wide range of $B_{\mathrm{ac}}$ (the sweep axis)
      and burst durations. The green and blue lines correspond respectively to the maxima of a cosine with and without a phase shift of
      $\pi/4$. 
      %In contrast to the green lines, the blue lines are located in advance of the maxima of the observed oscillations for small burst times, and behind the maxima for longer burst times.
       The current-to-field conversion factor $K$ is fitted for both cases separately ($K$=0.568 mT/mA and $K$=0.60 mT/mA for respectively with and without phase shift; the fit range is $t=60-500$ ns and $I_\mathrm{s}=3.6-6.3$ mA).
b) Phase shift for a wide range of $B_{\mathrm{ac}}$, displayed as a function of stripline current $I_\mathrm{s}$. Values obtained from a fit of each trace of the data in a) (varying burst time, constant $B_{\mathrm{ac}}$) to a damped cosine $a_1-a_2
\cos (\frac{1}{2}KI_\mathrm{s}g\mu_Bt+a_3\pi)/\sqrt{t}$, where $a_{1,2,3}$ are fit parameters and $K=0.568$ mT/mA. $I_\mathrm{s}$
is a known value in the experiment, extracted from the applied RF power. The gray
dashed lines
        represent the 95\% confidence-interval.}
\label{colorplot}
\end{figure}

We continue the discussion with the experimental observation of the second theoretically predicted prominent feature of the Rabi
oscillations, i.e., a phase shift of $\pi/4$ in the oscillations, which is independent of all parameters. 
%In principle, the
%experimental value of this phase shift $\phi$ can be extracted from single traces like those in Fig.
%\ref{oscillations}. However, the precision is poor because a small uncertainty in the Rabi frequency can lead to a large
%uncertainty in $\phi$. A much 
The value of $\phi$ can be extracted most accurately from the oscillations measured for a wide range and
small steps of $B_{\mathrm{ac}}$, like the data shown in Fig. \ref{colorplot}a. That is because the Rabi period $T_{Rabi}=2\pi/g\mu_B
(\frac{1}{2}B_{\mathrm{ac}})=2\pi/g\mu_B (\frac{1}{2}KI_\mathrm{s})$ contains only one unknown parameter $K$ (current to oscillating field amplitude $B_{\mathrm{ac}}$ conversion factor, in units of T/A)
which is independent of the current through the wire $I_\mathrm{s}$ that generates $B_{\mathrm{ac}}$ \cite{koppensnature}. The presence of a
phase shift is visible in Fig. \ref{colorplot}a, where the green and blue lines correspond respectively to the maxima of a cosine
with and without a phase shift of $\pi/4$. The green lines match very well the yellow bands representing high data values.
In contrast, the blue lines are located on the right side of the yellow bands for small burst times and more and more on the
left side of the bands for increasing burst times. Thus, a cosine without a phase shift does not match with the
observed Rabi oscillations.

In order to determine $\phi$ quantitatively, we perform a single two-dimensional fit of the complete dataset in Fig.
\ref{colorplot}a with $P_{odd}(t)$ (Eq. (\ref{eq:poddtwodot})), excluding the 1/t-term (see \cite{EPAPS}). The fit range is $t=100-900$ ns, such that the contribution from the 1/t-term of Eq. (\ref{eq:poddtwodot}) can be
neglected. For the 95\% confidence interval we find $\phi=(0.23\pm0.01)\pi$, close to the theoretical value. The relation between $\phi$ and $B_{\mathrm{ac}}$ is visible in Fig. \ref{colorplot}b, where we find no significant dependence of $\phi$ as a function of $B_{\mathrm{ac}}$, although the accuracy decreases for smaller $B_{\mathrm{ac}}$ (values obtained from fits to single traces, see caption). We have not compensated for the effects of the finite rise time ($<$2 ns) of the bursts, which leads to a small negative phase shift, on top of the expected positive $\pi/4$ shift.

To conclude, we have experimentally observed a power-law decay and universal phase shift of driven single electron spin oscillations. These features are theoretically understood by taking into account the coupling of the spin to the nuclear spin bath, which is static on the timescale of the electron spin evolution time. Furthermore, the slow power-law decay allows spin manipulation with relatively small driving fields. This improved understanding of the coherence of a driven single electron spin is important for future experiments using the electron spin as a qubit. For future investigation, it remains interesting to obtain more information about the non-static contributions of the nuclear bath or other
possible decoherence mechanisms.  For that, it is required to measure the driven oscillations at larger external fields, with
larger driving powers and longer evolution times than accessible in this work.

We thank T. Meunier, R. Hanson, Y.V. Nazarov and I.T. Vink for discussions; R. Schouten, A. van der Enden, R. Roeleveld and W.
den Braver for technical assistance. We acknowledge financial support from the Dutch Organization for Fundamental Research on
Matter (FOM), the Netherlands Organization for Scientific Research (NWO), JST ICORP, NCCR Nanoscience, and the Swiss NSF.

%\bibliographystyle{apsrev}
%\bibliography{Bib_koppens2}

\begin{thebibliography}{27}
\expandafter\ifx\csname natexlab\endcsname\relax\def\natexlab#1{#1}\fi
\expandafter\ifx\csname bibnamefont\endcsname\relax
  \def\bibnamefont#1{#1}\fi
\expandafter\ifx\csname bibfnamefont\endcsname\relax
  \def\bibfnamefont#1{#1}\fi
\expandafter\ifx\csname citenamefont\endcsname\relax
  \def\citenamefont#1{#1}\fi
\expandafter\ifx\csname url\endcsname\relax
  \def\url#1{\texttt{#1}}\fi
\expandafter\ifx\csname urlprefix\endcsname\relax\def\urlprefix{URL }\fi
\providecommand{\bibinfo}[2]{#2}
\providecommand{\eprint}[2][]{\url{#2}}

\bibitem[{\citenamefont{Bloch}(1946)}]{bloch1946}
\bibinfo{author}{\bibfnamefont{F.}~\bibnamefont{Bloch}},
  \bibinfo{journal}{Phys. Rev.} \textbf{\bibinfo{volume}{70}},
  \bibinfo{pages}{460} (\bibinfo{year}{1946}).

\bibitem[{\citenamefont{DiVincenzo and Loss}(2005)}]{lossprb05}
\bibinfo{author}{\bibfnamefont{D.~P.} \bibnamefont{DiVincenzo}}
  \bibnamefont{and} \bibinfo{author}{\bibfnamefont{D.}~\bibnamefont{Loss}},
  \bibinfo{journal}{Phys. Rev. B} \textbf{\bibinfo{volume}{71}},
  \bibinfo{pages}{035318} (\bibinfo{year}{2005}).

\bibitem[{\citenamefont{Taylor and Lukin}(2006)}]{taylor06}
\bibinfo{author}{\bibfnamefont{J.}~\bibnamefont{Taylor}} \bibnamefont{and}
  \bibinfo{author}{\bibfnamefont{M.}~\bibnamefont{Lukin}},
  \bibinfo{journal}{Quantum Information Processing}
  \textbf{\bibinfo{volume}{5}}, \bibinfo{pages}{503} (\bibinfo{year}{2006}).

\bibitem[{\citenamefont{Ithier et~al.}(2005)\citenamefont{Ithier, Collin,
  Joyez, Meeson, Vion, Esteve, Chiarello, Shnirman, Makhlin, Schriefl
  et~al.}}]{ithier05}
\bibinfo{author}{\bibfnamefont{G.}~\bibnamefont{Ithier \textit{et~al.}}},
  %\bibinfo{author}{\bibfnamefont{E.}~\bibnamefont{Collin}},
%  \bibinfo{author}{\bibfnamefont{P.}~\bibnamefont{Joyez}},
%  \bibinfo{author}{\bibfnamefont{P.}~\bibnamefont{Meeson}},
%  \bibinfo{author}{\bibfnamefont{D.}~\bibnamefont{Vion}},
%  \bibinfo{author}{\bibfnamefont{D.}~\bibnamefont{Esteve}},
%  \bibinfo{author}{\bibfnamefont{F.}~\bibnamefont{Chiarello}},
%  \bibinfo{author}{\bibfnamefont{A.}~\bibnamefont{Shnirman}},
%  \bibinfo{author}{\bibfnamefont{Y.}~\bibnamefont{Makhlin}},
%  \bibinfo{author}{\bibfnamefont{J.}~\bibnamefont{Schriefl}},
  \bibnamefont{et~al.}, \bibinfo{journal}{Physical Review B}
  \textbf{\bibinfo{volume}{72}}, \bibinfo{pages}{134519}
  (\bibinfo{year}{2005}).

\bibitem[{\citenamefont{Khaetskii and Nazarov}(2000)}]{khaetskiinazarov}
\bibinfo{author}{\bibfnamefont{A.V.}~\bibnamefont{Khaetskii}} \bibnamefont{and}
  \bibinfo{author}{\bibfnamefont{Y.V.}~\bibnamefont{Nazarov}},
  \bibinfo{journal}{Phys. Rev. B} \textbf{\bibinfo{volume}{61}},
  \bibinfo{pages}{12639} (\bibinfo{year}{2000}).

\bibitem[{\citenamefont{Golovach et~al.}(2004)\citenamefont{Golovach,
  Khaetskii, and Loss}}]{golovach03}
\bibinfo{author}{\bibfnamefont{V.~N.} \bibnamefont{Golovach}},
  \bibinfo{author}{\bibfnamefont{A.}~\bibnamefont{Khaetskii}},
  \bibnamefont{and} \bibinfo{author}{\bibfnamefont{D.}~\bibnamefont{Loss}},
  \bibinfo{journal}{\prl} \textbf{\bibinfo{volume}{93}},
  \bibinfo{pages}{016601} (\bibinfo{year}{2004}).

\bibitem[{\citenamefont{Elzerman et~al.}(2004)\citenamefont{Elzerman, Hanson,
  van Beveren, Witkamp, Vandersypen, and Kouwenhoven}}]{elzerman}
\bibinfo{author}{\bibfnamefont{J.~M.} \bibnamefont{Elzerman \textit{et~al.}}},
%  \bibinfo{author}{\bibfnamefont{R.}~\bibnamefont{Hanson}},
%  \bibinfo{author}{\bibfnamefont{L.~H.~W.} \bibnamefont{van Beveren}},
%  \bibinfo{author}{\bibfnamefont{B.}~\bibnamefont{Witkamp}},
%  \bibinfo{author}{\bibfnamefont{L.~M.~K.} \bibnamefont{Vandersypen}},
%  \bibnamefont{and} \bibinfo{author}{\bibfnamefont{L.~P.}
%  \bibnamefont{Kouwenhoven}}, 
\bibinfo{journal}{Nature}
  \textbf{\bibinfo{volume}{430}}, \bibinfo{pages}{431} (\bibinfo{year}{2004}).

\bibitem[{\citenamefont{Kroutvar et~al.}(2004)\citenamefont{Kroutvar, Ducommun,
  Heiss, Bichler, Schuh, Abstreiter, and Finley}}]{kroutvar}
\bibinfo{author}{\bibfnamefont{M.}~\bibnamefont{Kroutvar \textit{et~al.}}},
%  \bibinfo{author}{\bibfnamefont{Y.}~\bibnamefont{Ducommun}},
%  \bibinfo{author}{\bibfnamefont{D.}~\bibnamefont{Heiss}},
%  \bibinfo{author}{\bibfnamefont{M.}~\bibnamefont{Bichler}},
%  \bibinfo{author}{\bibfnamefont{D.}~\bibnamefont{Schuh}},
%  \bibinfo{author}{\bibfnamefont{G.}~\bibnamefont{Abstreiter}},
%  \bibnamefont{and} \bibinfo{author}{\bibfnamefont{J.~J.}
%  \bibnamefont{Finley}},
\bibinfo{journal}{Nature}
  \textbf{\bibinfo{volume}{432}}, \bibinfo{pages}{81} (\bibinfo{year}{2004}).

\bibitem[{\citenamefont{Amasha et~al.}(2006)\citenamefont{Amasha, MacLean,
  Radu, Zumbuhl, Kastner, Hanson, and Gossard}}]{amasha-2006}
\bibinfo{author}{\bibfnamefont{S.}~\bibnamefont{Amasha \textit{et~al.}}},
%  \bibinfo{author}{\bibfnamefont{K.}~\bibnamefont{MacLean}},
%  \bibinfo{author}{\bibfnamefont{I.}~\bibnamefont{Radu}},
%  \bibinfo{author}{\bibfnamefont{D.~M.} \bibnamefont{Zumbuhl}},
%  \bibinfo{author}{\bibfnamefont{M.~A.} \bibnamefont{Kastner}},
%  \bibinfo{author}{\bibfnamefont{M.~P.} \bibnamefont{Hanson}},
%  \bibnamefont{and} \bibinfo{author}{\bibfnamefont{A.~C.}
%  \bibnamefont{Gossard}},
\bibinfo{journal}{arXiv:cond-mat/0607110}
  (\bibinfo{year}{2006}).

\bibitem[{\citenamefont{Khaetskii et~al.}(2002)\citenamefont{Khaetskii, Loss,
  and Glazman}}]{khaetskii02}
\bibinfo{author}{\bibfnamefont{A.~V.} \bibnamefont{Khaetskii}},
  \bibinfo{author}{\bibfnamefont{D.}~\bibnamefont{Loss}}, \bibnamefont{and}
  \bibinfo{author}{\bibfnamefont{L.}~\bibnamefont{Glazman}},
  \bibinfo{journal}{\prl} \textbf{\bibinfo{volume}{88}},
  \bibinfo{pages}{186802} (\bibinfo{year}{2002}).

\bibitem[{mer()}]{merkulov}
\bibinfo{note}{I.A. Merkulov, A.L. Efros, M. Rosen, {Phys. Rev. B} {\bf 65},
  205309 (2002)}.

\bibitem[{\citenamefont{Coish and Loss}(2004)}]{coish04}
\bibinfo{author}{\bibfnamefont{W.~A.} \bibnamefont{Coish}} \bibnamefont{and}
  \bibinfo{author}{\bibfnamefont{D.}~\bibnamefont{Loss}},
  \bibinfo{journal}{\prb} \textbf{\bibinfo{volume}{70}},
  \bibinfo{pages}{195340} (\bibinfo{year}{2004}).

\bibitem[{\citenamefont{{Coish} and {Loss}}(2005)}]{coish05}
\bibinfo{author}{\bibfnamefont{W.~A.} \bibnamefont{{Coish}}} \bibnamefont{and}
  \bibinfo{author}{\bibfnamefont{D.}~\bibnamefont{{Loss}}},
  \bibinfo{journal}{Phys. Rev. B} \textbf{\bibinfo{volume}{72}},
  \bibinfo{pages}{125337} (\bibinfo{year}{2005}).

\bibitem[{\citenamefont{Johnson et~al.}(2005)\citenamefont{Johnson, Petta,
  Taylor, Yacoby, Lukin, Marcus, Hanson, and Gossard}}]{johnsonnature}
\bibinfo{author}{\bibfnamefont{A.~C.} \bibnamefont{Johnson \textit{et~al.}}},
%  \bibinfo{author}{\bibfnamefont{J.~R.} \bibnamefont{Petta}},
%  \bibinfo{author}{\bibfnamefont{J.~M.} \bibnamefont{Taylor}},
%  \bibinfo{author}{\bibfnamefont{A.}~\bibnamefont{Yacoby}},
%  \bibinfo{author}{\bibfnamefont{M.~D.} \bibnamefont{Lukin}},
%  \bibinfo{author}{\bibfnamefont{C.~M.} \bibnamefont{Marcus}},
%  \bibinfo{author}{\bibfnamefont{M.~P.} \bibnamefont{Hanson}},
%  \bibnamefont{and} \bibinfo{author}{\bibfnamefont{A.~C.}
%  \bibnamefont{Gossard}},
\bibinfo{journal}{Nature}
  \textbf{\bibinfo{volume}{435}}, \bibinfo{pages}{925} (\bibinfo{year}{2005}).

\bibitem[{\citenamefont{Koppens et~al.}(2005)\citenamefont{Koppens, Folk,
  Elzerman, Hanson, van Beveren, Vink, Tranitz, Wegscheider, Kouwenhoven, and
  Vandersypen}}]{koppensscience}
\bibinfo{author}{\bibfnamefont{F.~H.~L.} \bibnamefont{Koppens \textit{et~al.}}},
%  \bibinfo{author}{\bibfnamefont{J.~A.} \bibnamefont{Folk}},
%  \bibinfo{author}{\bibfnamefont{J.~M.} \bibnamefont{Elzerman}},
%  \bibinfo{author}{\bibfnamefont{R.}~\bibnamefont{Hanson}},
%  \bibinfo{author}{\bibfnamefont{L.~H.~W.} \bibnamefont{van Beveren}},
%  \bibinfo{author}{\bibfnamefont{I.~T.} \bibnamefont{Vink}},
%  \bibinfo{author}{\bibfnamefont{H.~P.} \bibnamefont{Tranitz}},
%  \bibinfo{author}{\bibfnamefont{W.}~\bibnamefont{Wegscheider}},
%  \bibinfo{author}{\bibfnamefont{L.~P.} \bibnamefont{Kouwenhoven}},
%  \bibnamefont{and} \bibinfo{author}{\bibfnamefont{L.~M.~K.}
%  \bibnamefont{Vandersypen}},
\bibinfo{journal}{Science}
  \textbf{\bibinfo{volume}{309}}, \bibinfo{pages}{1346} (\bibinfo{year}{2005}).

\bibitem[{\citenamefont{Petta et~al.}(2005)\citenamefont{Petta, Johnson,
  Taylor, Laird, Yacoby, Lukin, Marcus, Hanson, and Gossard}}]{petta}
\bibinfo{author}{\bibfnamefont{J.~R.} \bibnamefont{Petta \textit{et~al.}}},
%  \bibinfo{author}{\bibfnamefont{A.~C.} \bibnamefont{Johnson}},
%  \bibinfo{author}{\bibfnamefont{J.~M.} \bibnamefont{Taylor}},
%  \bibinfo{author}{\bibfnamefont{E.~A.} \bibnamefont{Laird}},
%  \bibinfo{author}{\bibfnamefont{A.}~\bibnamefont{Yacoby}},
%  \bibinfo{author}{\bibfnamefont{M.~D.} \bibnamefont{Lukin}},
%  \bibinfo{author}{\bibfnamefont{C.~M.} \bibnamefont{Marcus}},
%  \bibinfo{author}{\bibfnamefont{M.~P.} \bibnamefont{Hanson}},
%  \bibnamefont{and} \bibinfo{author}{\bibfnamefont{A.~C.}
%  \bibnamefont{Gossard}},
\bibinfo{journal}{Science}
  \textbf{\bibinfo{volume}{309}}, \bibinfo{pages}{2180} (\bibinfo{year}{2005}).

\bibitem[{\citenamefont{Laird et~al.}(2006)\citenamefont{Laird, Petta, Johnson,
  Marcus, Yacoby, Hanson, and Gossard}}]{laird06}
\bibinfo{author}{\bibfnamefont{E.~A.} \bibnamefont{Laird \textit{et~al.}}},
%  \bibinfo{author}{\bibfnamefont{J.~R.} \bibnamefont{Petta}},
%  \bibinfo{author}{\bibfnamefont{A.~C.} \bibnamefont{Johnson}},
%  \bibinfo{author}{\bibfnamefont{C.~M.} \bibnamefont{Marcus}},
%  \bibinfo{author}{\bibfnamefont{A.}~\bibnamefont{Yacoby}},
%  \bibinfo{author}{\bibfnamefont{M.~P.} \bibnamefont{Hanson}},
%  \bibnamefont{and} \bibinfo{author}{\bibfnamefont{A.~C.}
%  \bibnamefont{Gossard}},
\bibinfo{journal}{Phys. Rev. Lett.}
  \textbf{\bibinfo{volume}{97}}, \bibinfo{pages}{056801}
  (\bibinfo{year}{2006}).

\bibitem[{\citenamefont{Koppens et~al.}(2006)\citenamefont{Koppens, Buizert,
  Tielrooij, Vink, Nowack, Meunier, Kouwenhoven, and
  Vandersypen}}]{koppensnature}
\bibinfo{author}{\bibfnamefont{F.~H.~L.} \bibnamefont{Koppens \textit{et~al.}}},
%  \bibinfo{author}{\bibfnamefont{C.}~\bibnamefont{Buizert}},
%  \bibinfo{author}{\bibfnamefont{K.~J.} \bibnamefont{Tielrooij}},
%  \bibinfo{author}{\bibfnamefont{I.~T.} \bibnamefont{Vink}},
%  \bibinfo{author}{\bibfnamefont{K.~C.} \bibnamefont{Nowack}},
%  \bibinfo{author}{\bibfnamefont{T.}~\bibnamefont{Meunier}},
%  \bibinfo{author}{\bibfnamefont{L.~P.} \bibnamefont{Kouwenhoven}},
%  \bibnamefont{and} \bibinfo{author}{\bibfnamefont{L.~M.~K.}
%  \bibnamefont{Vandersypen}},
\bibinfo{journal}{Nature}
  \textbf{\bibinfo{volume}{442}}, \bibinfo{pages}{766} (\bibinfo{year}{2006}).

\bibitem[{\citenamefont{Yao et~al.}(2006)\citenamefont{Yao, Liu, and
  Sham}}]{yao}
\bibinfo{author}{\bibfnamefont{W.}~\bibnamefont{Yao}},
  \bibinfo{author}{\bibfnamefont{R.-B.} \bibnamefont{Liu}}, \bibnamefont{and}
  \bibinfo{author}{\bibfnamefont{L.~J.} \bibnamefont{Sham}},
  \bibinfo{journal}{Phys. Rev. B} \textbf{\bibinfo{volume}{74}},
  \bibinfo{pages}{195301} (\bibinfo{year}{2006}).

\bibitem[{\citenamefont{de~Sousa and Das~Sarma}(2003)}]{sousa03}
\bibinfo{author}{\bibfnamefont{R.}~\bibnamefont{de~Sousa}} \bibnamefont{and}
  \bibinfo{author}{\bibfnamefont{S.}~\bibnamefont{Das~Sarma}},
  \bibinfo{journal}{\prb} \textbf{\bibinfo{volume}{67}}, \bibinfo{pages}{33301}
  (\bibinfo{year}{2003}).

\bibitem[{\citenamefont{{Klauser} et~al.}(2006)\citenamefont{{Klauser},
  {Coish}, and {Loss}}}]{klauser06}
\bibinfo{author}{\bibfnamefont{D.}~\bibnamefont{{Klauser}}},
  \bibinfo{author}{\bibfnamefont{W.~A.} \bibnamefont{{Coish}}},
  \bibnamefont{and} \bibinfo{author}{\bibfnamefont{D.}~\bibnamefont{{Loss}}},
  \bibinfo{journal}{Phys. Rev. B} \textbf{\bibinfo{volume}{73}},
  \bibinfo{pages}{205302} (\bibinfo{year}{2006}).

\bibitem[{\citenamefont{Deng and Hu}(2006{\natexlab{a}})}]{deng}
\bibinfo{author}{\bibfnamefont{C.}~\bibnamefont{Deng}} \bibnamefont{and}
  \bibinfo{author}{\bibfnamefont{X.}~\bibnamefont{Hu}}, \bibinfo{journal}{Phys.
  Rev. B} \textbf{\bibinfo{volume}{73}}, \bibinfo{pages}{241303(R)}
  (\bibinfo{year}{2006}{\natexlab{a}}).

\bibitem[{\citenamefont{Deng and Hu}(2006{\natexlab{b}})}]{dengerratum}
\bibinfo{author}{\bibfnamefont{C.}~\bibnamefont{Deng}} \bibnamefont{and}
  \bibinfo{author}{\bibfnamefont{X.}~\bibnamefont{Hu}}, \bibinfo{journal}{Phys.
  Rev. B} \textbf{\bibinfo{volume}{74}}, \bibinfo{pages}{129902(E)}
  (\bibinfo{year}{2006}{\natexlab{b}}).

\bibitem[{EPA()}]{EPAPS}
\bibinfo{note}{See EPAPS Document No. [number will be inserted by publisher]
  for fit procedures and asymptotic expansion}.

\bibitem[{\citenamefont{Dobrovitski et~al.}(2003)\citenamefont{Dobrovitski,
  De~Raedt, Katsnelson, and Harmon}}]{harmon2003}
\bibinfo{author}{\bibfnamefont{V.~V.} \bibnamefont{Dobrovitski}},
  \bibinfo{author}{\bibfnamefont{H.~A.} \bibnamefont{De~Raedt}},
  \bibinfo{author}{\bibfnamefont{M.~I.} \bibnamefont{Katsnelson}},
  \bibnamefont{and} \bibinfo{author}{\bibfnamefont{B.~N.}
  \bibnamefont{Harmon}}, \bibinfo{journal}{Phys. Rev. Lett.}
  \textbf{\bibinfo{volume}{90}}, \bibinfo{pages}{210401}
  (\bibinfo{year}{2003}).

\bibitem[{\citenamefont{{Huang} et~al.}(2005)\citenamefont{{Huang},
  {Dobrovitski}, and {Harmon}}}]{huang:2005a}
\bibinfo{author}{\bibfnamefont{R.-S.} \bibnamefont{{Huang}}},
  \bibinfo{author}{\bibfnamefont{V.}~\bibnamefont{{Dobrovitski}}},
  \bibnamefont{and} \bibinfo{author}{\bibfnamefont{B.}~\bibnamefont{{Harmon}}},
  \bibinfo{journal}{arXiv:cond-mat/0504449}  (\bibinfo{year}{2005}).

\bibitem[{\citenamefont{Borhani et~al.}(2006)\citenamefont{Borhani, Golovach,
  and Loss}}]{borhani06}
\bibinfo{author}{\bibfnamefont{M.}~\bibnamefont{Borhani}},
  \bibinfo{author}{\bibfnamefont{V.~N.} \bibnamefont{Golovach}},
  \bibnamefont{and} \bibinfo{author}{\bibfnamefont{D.}~\bibnamefont{Loss}},
  \bibinfo{journal}{Phys. Rev. B} \textbf{\bibinfo{volume}{73}},
  \bibinfo{pages}{155311} (\bibinfo{year}{2006}).

\end{thebibliography}
\newcommand{\noopsort}[1]{} \newcommand{\printfirst}[2]{#1}
  \newcommand{\singleletter}[1]{#1} \newcommand{\switchargs}[2]{#2#1}

\newpage
\cleardoublepage
\setcounter{figure}{0}
\setcounter{equation}{0}
\section{Supplementary material for ``Universal phase shift and
  non-exponential decay of driven single-spin oscillations''}

\subsection{Fit Procedures}

\begin{figure}[h]
          \includegraphics[scale=1]{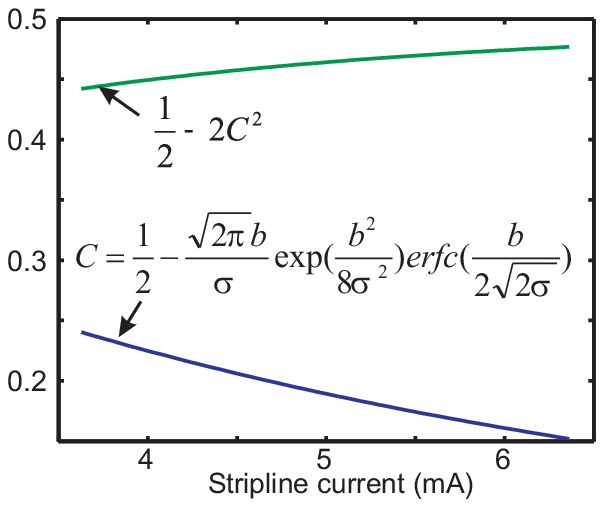}
   %  \centering
     \caption{ $C$ and $0.5-2C^2$ as a function of $I_s$, with $\sigma=g\mu_B(1.4$ mT)
and $b=g \mu_B K I_s$. $K=0.56$ mT/mA for the curves shown here, but the curves remain linear as well for $K$=0.5-0.6 mT/mA.}
     \label{C}
\end{figure}

\begin{figure} [h]
          \includegraphics[scale=1]{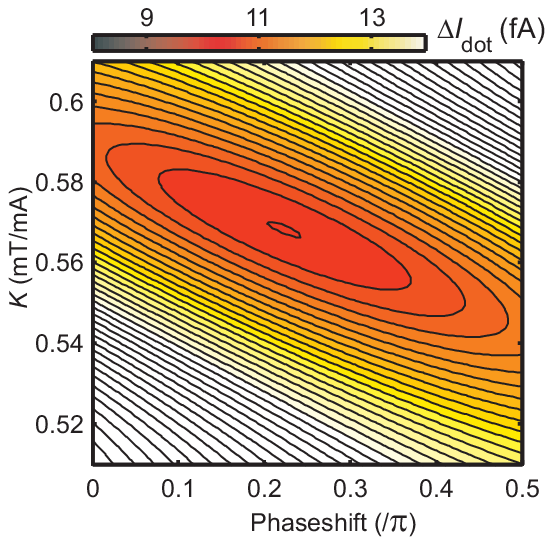}
   %  \centering
     \caption{ Phase shift and  fit quality for a range of current-to-field conversion factors $K$. Values obtained from a fit with a damped cosine to the single traces (constant $I_s$) of the data set shown in
     Fig. 3a of the main text, and subsequently averaged over all traces for the range $t=100-900$ ns and $I_s=$3.6-6.3 mA.
     The fit-quality $R^2$ is a measure of the correlation between the observed values $y_i$ and the predicted values
     ${\hat{y}}_i$: $R^2=\sum_{i=1}^n\frac{({\hat{y}}_i-{\bar{y}}_i)^2}{(y_i-{\bar{y}}_i)^2}$, with $\bar{y}=\frac{1}{n}\sum{y_i}$.}
     \label{phaseshift}
\end{figure}

Here, we describe the exact procedure of the two-dimensional fit from which the phase shift $\phi=(0.23\pm0.01)$ was obtained.
The fit function is a simplification of $P_\mathrm{odd}(t)$ (Eq. (4)). The first simplification is the exclusion of the 1/t-term because
its contribution is negligible within the fit range of $t=100-900$ ns. Second, both expressions for $C$ and $0.5-2C^2$ are
approximated as being linear in $b$, which is justified for the regime we consider ($I_s=3.6-6.3$ mA), as can be seen in Fig.
\ref{C}. With these simplifications, we obtain the expression $y_\mathrm{model}=a_1+I_sa_2+(a_3+I_sa_4)cos(\frac{1}{2}KI_sg\mu_Bt/\hbar+\phi
\pi)/\sqrt{t}$, where $a_{1,2,3,4}$, $K$ and $\phi$ are fit parameters. The Rabi frequency $\omega_\mathrm{Rabi}$ is given by $\frac{1}{2}KI_sg\mu_B/\hbar$,
with $I_s$ the current through the stripline which is known in the experiment. The constant factor $K=B_{ac}/I_s$ is not known in
the experiment but can be obtained from the fit. The behavior of $y_\mathrm{model}-y_\mathrm{data}$ around the optimal values for the fit parameters is seen in Fig. \ref{phaseshift}.

\begin{figure}[h]
          \includegraphics[scale=1]{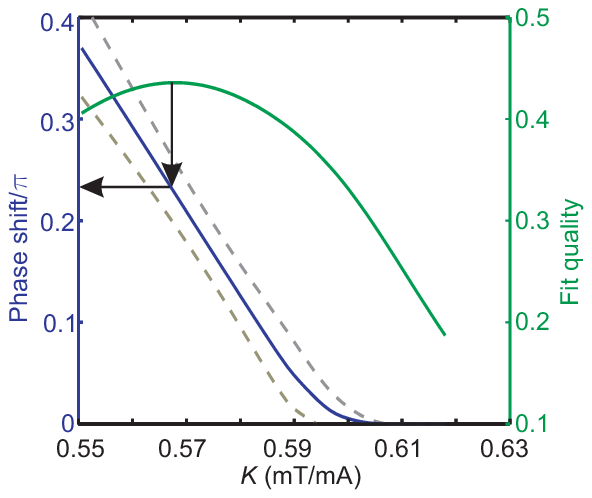}
   %  \centering
     \caption{ Root mean square difference between the measured current $y_\mathrm{data}$ and the model $y_\mathrm{model}$: $\left(\displaystyle\sum_{t,I_s}{(y^\mathrm{data}_{t,I_s}-y^\mathrm{model}_{t,I_s})^2}\right)^\frac{1}{2}/N_tN_{I_s}$, for a wide range
     of $K$ and $\phi$. We sum over the range $t=100-900$ ns and $I_s=$3.6-6.3 mA.}
     \label{residual}
\end{figure}

As a cross-check, we carried out a fit for each trace of the data in Fig.\,3a of the main text (varying burst time,
constant $B_{\mathrm{ac}}$) with a damped cosine $a_1-a_2\cos(\frac{1}{2}KI_sg\mu_Bt/\hbar+\phi \pi)/\sqrt{t}$, where $a_{1,2}$ and $\phi$ are fit
parameters, and $K$ is kept at a constant value. This fit is carried out for a wide range of values for $K$ and $I_s$. The
best-fit values for $\phi$ obtained for different $I_s$ are averaged and plotted as a function of $K$ in Fig. \ref{residual}.
(blue curve), together with the spread in $\phi$ (gray dotted lines) and the fit quality (green curve). The figure shows that the best fit is obtained for $\phi=0.23\pi$, the same value we found from the two-dimensional fit.

\subsection{Asymptotic expansion}

Here we give steps and additional justification leading to the asymptotic expansion given in
Eq. (3) of the main text.
We consider averaging Eq. (2) from the main text over a quasicontinuous
distribution of $h_z$ values for the case where $\delta_{\omega}=h_z-h_0$ (with the replacement $\delta_{\omega}\rightarrow x$):

\begin{equation}
P_\uparrow(t)  =  \int_{-\infty}^{\infty }dx D(x) P_{\uparrow,x}(t).\label{eq:PupDefinition}
\end{equation}
As a consequence of the central-limit theorem, for a large number of nuclear spins in a random unpolarized state, the
distribution function $D(x)$ is well-approximated by a Gaussian with standard deviation $\sigma$
centered at $x=0$:
\begin{equation}
D(x)=\frac{1}{\sqrt{2\pi}\sigma}e^{-\frac{x^2}{2\sigma^2}}.\label{eq:GaussianDistribution}
\end{equation}
Inserting Eq. (\ref{eq:GaussianDistribution}) into
Eq. (\ref{eq:PupDefinition}) gives the sum of a time-independent part, which can be
evaluated exactly, and a time-dependent interference term $I(t)$:
\begin{equation}
P_\uparrow(t)=\frac{1}{2}+C+I(t).
\end{equation}
Here, C is given following Eq. (3) of the main text. With the change of variables $u=\left(\sqrt{b^2+4 x^2}-b\right)/2\sigma$,
and using the fact that the integrand is an even function of $x$, the
interference term becomes $I(t)=\mathrm{Re}\tilde{I}(t)$, where
\begin{equation}
\tilde{I}(t)=\sqrt{\frac{b}{8\pi\sigma}}e^{i bt/2}\int_0^\infty
du\frac{\exp\left(-\frac{u^2}{2}-\frac{bu}{2\sigma}+i\sigma t
    u\right)}{\sqrt{u}\sqrt{1+\frac{\sigma u}{b}}\left(1+\frac{2\sigma u}{b}\right)}.
\end{equation}
When $\sigma t \gg 1$, the time dependence of $\tilde{I}(t)$ is controlled by
the region $u\lesssim 1/\sigma t$.  The integrand simplifies considerably
for $\sigma u/b\ll 1$, which coincides with $\sigma t\gg 1$ for
\begin{equation}
u\lesssim\frac{1}{\sigma t}\ll \frac{b}{\sigma},\; t\gg \frac{1}{\sigma}.
\end{equation}
Equivalently, for
\begin{equation}
t>\max\left(\frac{1}{b},\frac{1}{\sigma}\right),
\end{equation}
we expand the integrand for $u< \min(1,b/\sigma)$:
\begin{widetext}
\begin{equation}
\tilde{I}(t) =
\sqrt{\frac{b}{8\pi\sigma}}e^{ibt/2}\int_0^\infty du\frac{\exp\left(-\lambda
    u+\mathcal{O}(u^2)\right)}{\sqrt{u}}\left(1+\mathcal{O}\left(\frac{\sigma u}{b}\right)\right),\;\;\;\lambda=\frac{b}{2\sigma}-i\sigma t. \label{eq:IntegrandExpand}
\end{equation}
\end{widetext}
Neglecting corrections of order $u^2$ in the exponential and order $\sigma
u/b$ in the integrand prefactor, the remaining integral can be
evaluated easily, giving
\begin{eqnarray}
I(t)&\sim&\frac{\cos\left[bt/2+\arctan(t/\tau)/2\right]}{2\left[1+\left(t/\tau\right)^2\right]^{1/4}},
\tau = b/2\sigma^2,\label{eq:IAsymp}\\
t &>&\max\left(1/b,1/\sigma\right).
\end{eqnarray}
Eq. (\ref{eq:IAsymp}) is valid for the time scale indicated for an
arbitrary ratio of $b/2\sigma$.  Due to the exponential cutoff at
$u\lesssim 2\sigma/b$ in Eq. (\ref{eq:IntegrandExpand}),
Eq. (\ref{eq:IAsymp}) is actually valid for all times in the limit
$b/2\sigma\gg 1$.  Expanding Eq. (\ref{eq:IAsymp}) to leading order for
$t/\tau\gg1$ gives the result in Eq. (3) of the main text.  Higher-order
contributions to the long-time expansion of Eq. (\ref{eq:IAsymp}) and
contributions due to corrections of order $\sigma u/b$ in
Eq. (\ref{eq:IntegrandExpand}) both lead to more rapidly decaying
behavior of order $\sim1/t^{3/2}$. The reason for the different phase shift here ($\pi/4$) relative to that found in Ref. \cite{coish05} ($3\pi/4$) is that here the fluctuations are \emph{longitudinal}, while in Ref. \cite{coish05} the fluctuations are along the transverse direction. This leads to a different integrand in Eq. (S4) and thus to a different value for the phase shift and decay power.

Since the long-time behavior of $I(t)$ is dominated by the form of the
integrand near $x=0$, the same result can be found after replacing
the Gaussian distribution function by any other distribution function
$\tilde{D}(x)$ which is analytic and has a single peak at $x=0$.  Specifically,
\begin{eqnarray}
\tilde{D}(x)&=&\tilde{D}(0)\exp\{-\frac{x^2}{2\sigma^2}+\mathcal{O}(x^3)\},\\
\frac{1}{\sigma^2}&=&-\left.\frac{d^2\ln \tilde{D}(x)}{dx^2}\right|_{x=0}.
\end{eqnarray}
Thus, the universal form of the long-time power-law decay and phase shift are relatively insensitive to the specific shape of the distribution
function.

\end{document}